%% file: hadron2011.tex
\documentclass[a4paper,11pt]{article}

\usepackage{contribution}

\input{econfmacros}
\input{contributionmacros}

\begin{document}

\input{contribution}

\end{document}

%% file: econfmacros.tex
\newcommand{\weblink}[2][]{%
    \ifthenelse{\equal{#1}{}}%
    {\textnormal{\url{#2}}}%
    {\textnormal{\href{#2}{#1}}}%
}

\newcommand{\acknowledgements}[1]{%
  \bigskip\bigskip
  \textsf{\textbf{\Large Acknowledgements}} \\[2ex]
  {#1}
  \bigskip
}

\def\beq{\begin{equation}}
\def\eeq#1{\label{#1}\end{equation}}
\def\eeqn{\end{equation}}

\def\beqa{\begin{eqnarray}}
\def\eeqa#1{\label{#1}\end{eqnarray}}
\def\eeqan{\end{eqnarray}}

\let\bar=\overbar

\def\Dslash{\not{\hbox{\kern-4pt $D$}}}
\def\dslash{\not{\hbox{\kern-2pt $\del$}}}

\def\msb{{\bar{\ssstyle M \kern -1pt S}}}

%% file: contributionmacros.tex
\newcommand{\contribution}[7][]{%
  \clearpage
  \thispagestyle{plain}
  \ifthenelse{\equal{#1}{}}
  {\hypersetup{pdftitle={#2}}}
  {\hypersetup{pdftitle={#1}}}
  \hypersetup{pdfauthor={{#3} {#4}}}
  {\centering\normalfont\LARGE\bfseries\sffamily #2 \par\nobreak}
  \lhead{}
  \chead{%
    \textit{\footnotesize XIV International Conference on Hadron Spectroscopy
      (\weblink[\textit{hadron2011}]{http://www.hadron2011.de}), 13-17 June 2011, Munich, Germany}%
  }
  \rhead{}
  \bigskip
  \begin{center}
    {#3} {#4}\ifthenelse{\equal{#6}{}}{}{\footnote{\weblink[#6]{mailto:#6}}}
    \ifthenelse{\equal{#7}{}}{}{#7} \\
    \textit{#5}
  \end{center}
  \bigskip
}

\renewcommand{\abstract}[1]{%
  \begin{center}
    \begin{minipage}{0.85\textwidth}
      \begin{footnotesize}
        #1
      \end{footnotesize}
    \end{minipage}
  \end{center}
  \bigskip
}

%% file: contribution.tex
%
%
%
%
%
{  



\title{On lepton pair production in proton-antiproton collisions at intermediate energies and the main backgrounds. }

\author{  A.N.~Skachkova,\\
Joint Institute for Nuclear Research, Dubna, Russia\\
E-mail:   Anna.Skachkova@cern.ch}
\date{}
\maketitle

\contribution[On lepton pair production in p-pbar collisions]  
{On lepton pair production in proton-antiproton collisions at intermediate energies and the main backgrounds.}  
{Anna}{Skachkova}  
{Joint Institute for Nuclear Research \\ 
 Joliot-Curie 6, 141980 Dubna, Moscow region, Russia}  
{Anna.Skachkova@cern.ch}  
{on behalf of the PANDA Collaboration}  
%

\abstract{
The lepton pair production via the quark-antiquark annihilation subprocess in collisions of beam antiproton with the proton target at $E_{beam}$ = 14 GeV (which corresponds to the center-of-mass energy of the $p \bar p$ system $E_{cm}$ = 5.3 GeV) is studied on the basis of the event sample simulated by PYTHIA6 generator and PandaRoot package. Different kinematical variables which may be useful for  design of the muon system and the electromagnetic calorimeter of the detector of PANDA experiment at FAIR, as well as for the study of proton structure functions in the available $x-Q^2$ kinematical region, are considered. 
The problems due to the presence of fake leptons that appear from meson decays, as well as due to the contribution of background QCD processes and minimum bias events, are also discussed. The set of cuts which allows one to separate the events with the signal lepton pairs from different kind of background events
are proposed.
}
%

\section{Introduction}

This intermediate energy experiment ($E_{beam}$ < 15 GeV) may play an important role because it allows to study the energy range where the perturbative methods of QCD (pQCD) come into interplay with a rich physics of bound states and resonances. 
A detailed and high-precision experimental study at PANDA may allow to discriminate between a large variety of existing nonperturbative approaches and models that already exist or are under development now.
Dilepton events may serve as a powerful tool to get out the information about the parton distribution functions (PDFs) in hadrons \cite{ARXIV}. The plans to study this process are included into the LoI and TPR of PANDA experiment at HESR. This study may provide an interesting information about quark dynamics inside the nucleon \cite{ECHAJA}.
The results of study of leptons angle and energy spectra distributions, based on this Monte-Carlo simulation, was used for a proper geometrical design of PANDA muon system.

\section{Observations and Interpretation}


The work presents the distributions of the most  essential kinematical
variables of individual leptons from $\bar{p}p \rightarrow l^{+}l^{-} + X$ (MMT-DY) and benchmark
$\bar{p}p \rightarrow  J/\Psi + X$ (J/$\Psi \rightarrow l^{+}l^{-}$) processes.
%
These distributions allow one to estimate the energy, transverse momentum and
angle ranges that may be covered by leptons produced in quark-antiquark annihilation process.
The PYTHIA6 simulation has shown that one may expect to gain
   about $7 \cdot 10^{7}$  MMT-DY events per year for the
   luminosity $L = 2 \cdot 10^{5} mb^{-1}s^{-1}$.


  The study of kinematical characteristics
   of lepton pair  as a whole system was also done. 
%
%
%
  The analysis of distributions
allowed to determine  the region in
   x-Q$^{2}$-plane which can be available for measuring 
    the  proton  structure function at PANDA: 
   $0.05 \leq x \leq 0.7$ and  $Q^{2} \leq 6.25$ GeV.
        
    An important problem  of  background suppresion
 is also considered.
The histograms which demonstrate the relative contribution of different
  parents and grandparents  of produced leptons
 are presented.
%
%
 According to PYTHIA, the fraction of signal dimuon events which include fake
  muons  is about 16.6$\%$.
In  a case of   electrons
the number of signal events fraction
 containing fake electrons is about 2$\%$.
   The set of three cuts is proposed 
   which
 allows to  reduce the fraction
   of the signal events containig fake decay leptons to the values
$fr_{\mu}=0001\%$ in   a case of  $\mu^{+}\mu^{-}$ production and
   $fr_{e}=0.008\%$ in $e^{+}e^{-}$. 
%
   Much more dangerous background is caused  by  minimum-bias 
  and QCD events.
The proposed set of five cuts allows 
  to get rid completely of minimum-bias and QCD background
  contribution in the $\mu^{+}\mu^{-}$ case  and to 
  reach the value of S/B =  3.8 for the $e^{+}e^{-}$  case.

 It was also  noted that the  study of events 
  with two (and even three) lepton pairs
  would allow to improve the precision of the
  parameters of multiple quark interactions, which measurement 
  will extend the
  region of QCD studies.
\acknowledgements{%
I am grateful to my coauthors for useful discussions and proposed topics for  investigation. Additional thanks to  FAIR-Russia Research Center (together with Federal Agency for Atomic Energy (Rosatom) and Helmholtz Association) for finantial support.
}

\vskip -2.5cm

%

}  
